\documentclass[reprint,aip,floatfix,apl]{revtex4-1}

\usepackage[pdftex]{graphicx}
\usepackage{dcolumn}
\usepackage{bm}
\usepackage{color}

\begin{document}

\title{Global consequences of a local Casimir force: Adhered cantilever}

\author{V. B. Svetovoy}
\email[Corresponding author: ]{v.b.svetovoy@rug.nl}
\affiliation{Zernike Institute for Advanced Materials, University of Groningen - Nijenborgh 4, 9747 AG Groningen, The Netherlands}
\affiliation{Yaroslavl Branch of the Institute of Physics and Technology, Russian Academy of Sciencies, Universitetskaya 21, 150007 Yaroslavl, Russia}
\author{A. E. Melenev}
\affiliation{P. G. Demidov Yaroslavl State University, Sovetskaya 14, 150000 Yaroslavl, Russia}
\author{M. V. Lokhanin}
\affiliation{P. G. Demidov Yaroslavl State University, Sovetskaya 14, 150000 Yaroslavl, Russia}
\author{G. Palasantzas}
\affiliation{Zernike Institute for Advanced Materials, University of Groningen - Nijenborgh 4, 9747 AG Groningen, The Netherlands}

\begin{abstract}
Although stiction is a cumbersome problem for microsystems, it stimulates investigations of surface adhesion. In fact, the shape of an adhered cantilever carries information of the adhesion energy that locks one end to the substrate. We demonstrate here that the system is also sensitive to the dispersion forces that are operative very close to the point of contact, but their contribution to the shape is maximum at about one third of the unadhered length. When the force exceeds a critical value the cantilever does not lose stability but it settles at smaller unadhered length, whose relation to adhesion energy is only slightly affected by the force. Our calculations suggest to use adhered cantilevers to measure the dispersion forces at short separations, where other methods suffer from jump-to-contact instability. Simultaneous measurement of the force and adhesion energy allows the separation of the dispersion contribution to the surface adhesion.
\end{abstract}

\maketitle

The dispersion forces, a common name for the fluctuation-induced van der Waals (vdW) and Casimir forces \cite{Dzyaloshinskii1961,Mahanty1976}, become measurable with relative ease at separations less than $100\:$nm \cite{Harris2000,Chan2001,Decca2003,Zwol2008} since they have significant magnitude. However, even at these separations they are weak compared to background forces such as elastic, electrostatic, or capillary forces. Only at very small separations between bodies $\sim 1\:$nm do the dispersion forces dominate. The latter means that these forces play a crucial role only near or at the point of contact of two macroscopic bodies. Although it is natural to expect that these forces are not important far away from the point of contact as, for example, was formulated in the crack theory by Barenblatt \cite{Barenblatt1962}, there are physical situations where the finite range of the dispersion interaction plays a principal role.

One example that was considered recently \cite{Svetovoy2016} demonstrated this effect for surface nanobubbles. These nanobubbles are gaseous domains trapped at the solid-liquid interface \cite{Lohse2015,Alheshibri2016}. They have the shape of a spherical cap with heights of $\sim 10\:$nm. The liquid and solid separated by a gaseous gap attract each other due to the dispersion interaction. The energy associated with this interaction at distances $d\sim 10\:$nm is estimated as $\sim 10^{-5}\:$J/m$^2$ that is much smaller than the surface tension of liquids $\gamma\sim 10^{-2}\:$J/m$^2$. However, in the corners, where the gas-solid and gas-liquid interfaces are met, the energy is singular. The singularity is resolved due to balance of the attractive vdW and repulsive chemical interaction at distances $\sim 3\:${\AA} \cite{Israelachvili2011}. For a drop in gas or in another liquid the effect of the dispersion interaction is important only at the very corners \cite{Getta1998}. However, for nanobubbles both the gas compressibility and a finite range of interaction influence significantly the global characteristics of the bubbles such as the aspect ratio or the contact angle \cite{Svetovoy2016}.

\begin{figure}[thb]
\centering
\includegraphics[width=86mm]{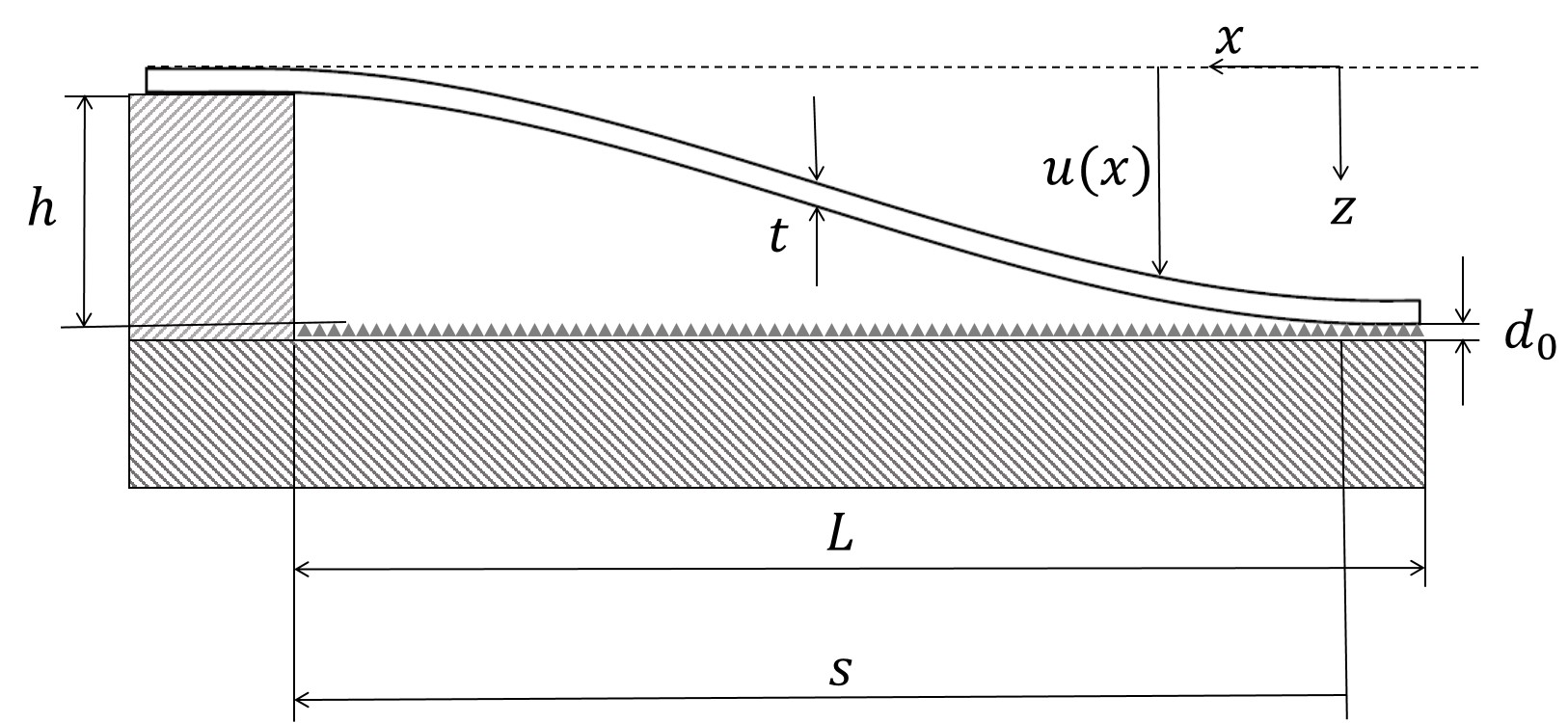}
\caption{Adhered cantilever. One end is firmly fixed at a height $h+d_0$ above the substrate. The other end sticks to the substrate. The coordinate system is chosen as shown in the figure. The cantilever and the substrate both are rough. A combined roughness defines the minimal distance at the contact $d_0$. The function $z=u(x)$ describes the shape of the cantilever in the coordinates $x-z$.   \label{fig:scheme}}
\end{figure}

Furthermore, the dispersion interaction close to the point of contact can influence the global characteristics of contacting bodies, which is a crucial issue in the fabrication and operation of micro/nano devices and architectures. The basic system under consideration is an adhered cantilever shown in Fig.$\:$\ref{fig:scheme}. The adhered cantilever problem originates from microfabrication, where unwanted stiction can appear during the final fabrication step (drying) or as an accidental stiction during operation  \cite{Maboudian1997,Parker2005}. A relation between the adhesion energy per unit area $\Gamma$ and the length of the not adhered part of the cantilever (crack) $s$ was established \cite{Mastrangelo1993a,Mastrangelo1993b,Boer1999}. This relation was used to measure the adhesion energy \cite{Knapp2002}. It was found \cite{DelRio2005} that in dry conditions the main contribution to $\Gamma$ comes from the dispersion interaction at the contact distance $d_0$. However, from the analysis of a restricted range of parameters it was concluded that the same dispersion interaction outside of the contact range gives only a small correction to the shape of cantilever \cite{Knapp2002}.

The Casimir force is typically measured between two bodies when one of them is suspended on an elastic spring \cite{Capasso2007}. At small distances the system becomes unstable and jumps to contact \cite{Barsenas2005,Esquivel2009,Broer2015}. Due to this instability the measurement of the force in air of vacuum at distances of the order or below $10\:$nm becomes cumbersome \cite{Palasantzas2008,Zwol2008,Tonck1991}. The adhered cantilever is an interesting system that can, in principle, overcome the problem of instability at small distances. This cantilever is never in an unstable state: if the force increases the crack length becomes smaller but the system does not lose stability. Therefore, if the dispersion forces can give a measurable contribution to the shape of the cantilever, we can extract information on the forces at distances, which are not available for the elastic suspension method, namely, below $13\:$nm \cite{Palasantzas2008}.

In this paper we analyse the influence of the vdW/Casimir force on the shape of the adhered cantilever and demonstrate that the effect of the force is measurable. Moreover, the contribution to the shape of the cantilever is maximal far from the point of contact where it is convenient to measure this contribution.

The system under consideration, the choice of the coordinate system, and the corresponding parameters are shown in Fig.$\:$\ref{fig:scheme} (note that the $x$-direction is positive to the left). The beam of length $L$, width $w$, and thickness $t$ is adhered to the substrate at the minimal distance $d_0$. The latter is defined by the combined roughness of the bodies in contact and is determined not by the root-mean-square roughness but by the highest asperities. A part of the beam of length $L-s$ sticks to the substrate with the adhesion energy per unit area $\Gamma$. The left end of the beam is firmly fixed at a height $h+d_0$ above the substrate. Homogeneous situation is assumed along the beam width ($y$-direction). The main objective is to find the shape of the beam $u(x)$ including the dispersion forces acting between the beam and the substrate outside of the contact area.

The total energy of the system can be presented in the form $E_{tot}=U(s)-\Gamma w (L-s)$, where $U(s)$ is the energy of the deformed part of the beam and the second term is the surface energy. Minimization of the total energy on $s$ gives the relation between $U(s)$ and the adhesion energy \cite{Boer1999}: $\Gamma =-w^{-1}dU/ds$. On the other hand, $U$ can be presented as a functional of $u(x)$
\begin{equation}\label{eq:energy}
  U[u]=w\int_0^s dx\left[\frac{D}{2}\left(\frac{d^2u}{dx^2}\right)^2-
  \int_0^uP(x,v)dv\right],
\end{equation}
where $x$ is the coordinate along the beam, $D=Et^3/12$ is the flexural rigidity, and $E$ is the Young's modulus of the beam material. The first term here is the elastic energy, while the second one is the work done by the external force per unit area $P(x,u)$.

Minimization of the functional $U[u]$ gives an equation for the beam shape
\begin{equation}\label{eq:beam}
  D\frac{d^{4}u}{dx^4}=P(x,u),
\end{equation}
which has to be solved with boundary conditions $u(0)=h$, $u^{\prime}(0)=0$, $u(s)=u^{\prime}(s)=0$, where the prime means a derivative with respect to the argument. We consider the case when $P(x,u)$ is the vdW/Casimir pressure. This pressure behaves with the separation gap $d$ as $ d^{-\alpha}$, where the exponent $3<\alpha<4$ is a weak function of $d$ and the local gap is $d=h-u(x)$. In a restricted range of separation distances $\alpha$ can be considered as a constant. Such a pressure can be presented in the form
\begin{equation}\label{eq:force}
  P(x,u)=P_C\left(1+R-R\zeta\right)^{-\alpha},\ R=h/d_0,\ \zeta=u/h,
\end{equation}
Here $1-\zeta$ is the normalized gap, $P_C$ is the pressure at $d_0$ and the parameter $R\gg 1$ is always large. At $\alpha=3$ or 4 we have pure vdW or pure Casimir pressures, respectively, but at separation distances of interest the interaction is in the transition region between the retarded and nonretarded cases and $\alpha$ is in between 3 and 4. Although the beam is curved we use the force between parallel plates that is well justified since the curvature radius of the cantilever is much larger than any other length scale.

Introducing the normalized coordinate $\xi$ and the force parameter $K$ the problem becomes completely dimensionless:
\begin{eqnarray}\label{eq:beam_less}
\nonumber \frac{d^{4}\zeta}{d\xi^4}=K^4\left(1+R-R\zeta\right)^{-\alpha},\\
&&\lefteqn{\hspace{-5.5cm}\xi=x/s,\ K=(P_C/P_s)^{1/4},\ P_s=Et^3h/12s^4.}
\end{eqnarray}
Here $P_s$ defines a pressure scale related to the elastic properties of the beam and $K^4$ is the relative measure of the dispersion pressure. Equation (\ref{eq:beam_less}) is a nonlinear boundary problem that has to be solved with the conditions: $\zeta(0)=1$, $\zeta^{\prime}(0)=0$, $\zeta(1)=0$, and $\zeta^{\prime}(1)=0$. The numerical solution of the problem is straightforward and can be found by a simple shooting method choosing $\zeta^{\prime\prime}$ and $\zeta^{\prime\prime\prime}$ at $\xi=0$ to satisfy the boundary conditions at $\xi=1$.

\begin{figure}[thb]
\centering
\includegraphics[width=86mm]{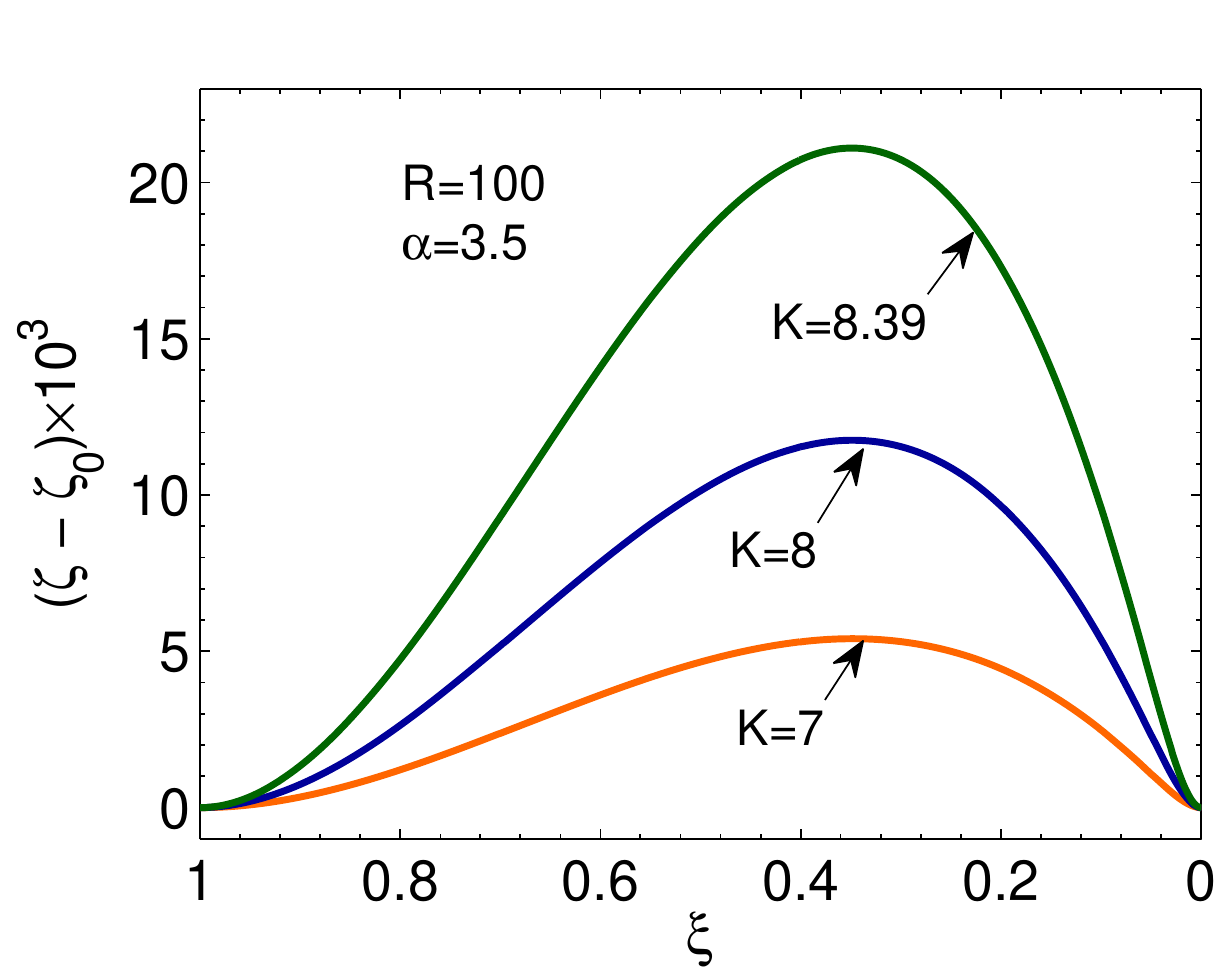}
\caption{Contribution of the dispersion pressure to the shape of the cantilever for three different values of $K$. There is a critical parameter $K_c$ such that for $K>K_c$ the solution disappears.  \label{fig:dz_num}}
\end{figure}

If the dispersion pressure is zero, $K=0$, an analytical solution exists that is
\begin{equation}\label{eq:zeta_0}
  \zeta_0=1-3\xi^2+2\xi^3.
\end{equation}
It describes the unperturbed shape of the beam shown in Fig.$\:$\ref{fig:scheme}. For a few nonzero values of $K$ the numerical solutions are presented in Fig.$\:$\ref{fig:dz_num}. It is interesting to note that the maximal deviation from the unperturbed shape $\zeta_0$ happens far from the point of contact at $\xi\approx 1/3$. On the other hand, the dispersion pressure decreases roughly one order of magnitude near the point of contact at the lateral distance $\xi_0\sim 1/\sqrt{3R}\ll 1$. Such a nonlocal response of the beam on the well localized dispersion pressure is explained by the boundary conditions at $\xi=0$ that do not allow fast changes of $\zeta$ near the point of contact.

The magnitude of the effect is unexpectedly large. For example, for $K=8$ the normalized deviation from the unperturbed shape of the cantilever is $0.011$. For the height $h=2\:\mu$m as in \cite{Knapp2002} we find that the deviation in absolute units is $22\:$nm. It can be compared with the $3\:$nm found in \cite{Knapp2002}, which corresponds to a significantly smaller value of $K$.

When $K$ becomes larger than some critical value the solution disappears. It happens at $K=K_c$ where the critical value $K_c$ at $R=100$ and $\alpha=3.5$ is $K_c\approx 8.39$. The deviation $\zeta-\zeta_0$ corresponding to this critical case is the largest. The solution disappears due to the following reason. At $K=0$ one has $\zeta^{\prime\prime}(0)=-6$ but when $K$ increases $\zeta^{\prime\prime}(0)$ increases too and becomes zero at $K=K_c$. The second derivative has to stay nonpositive at the point of contact otherwise the beam has to go below the substrate. Physically it means that a strong enough force is able to heal the crack reducing the length $s$ of the unadhered part.

\begin{figure}[thb]
\centering
\includegraphics[width=86mm]{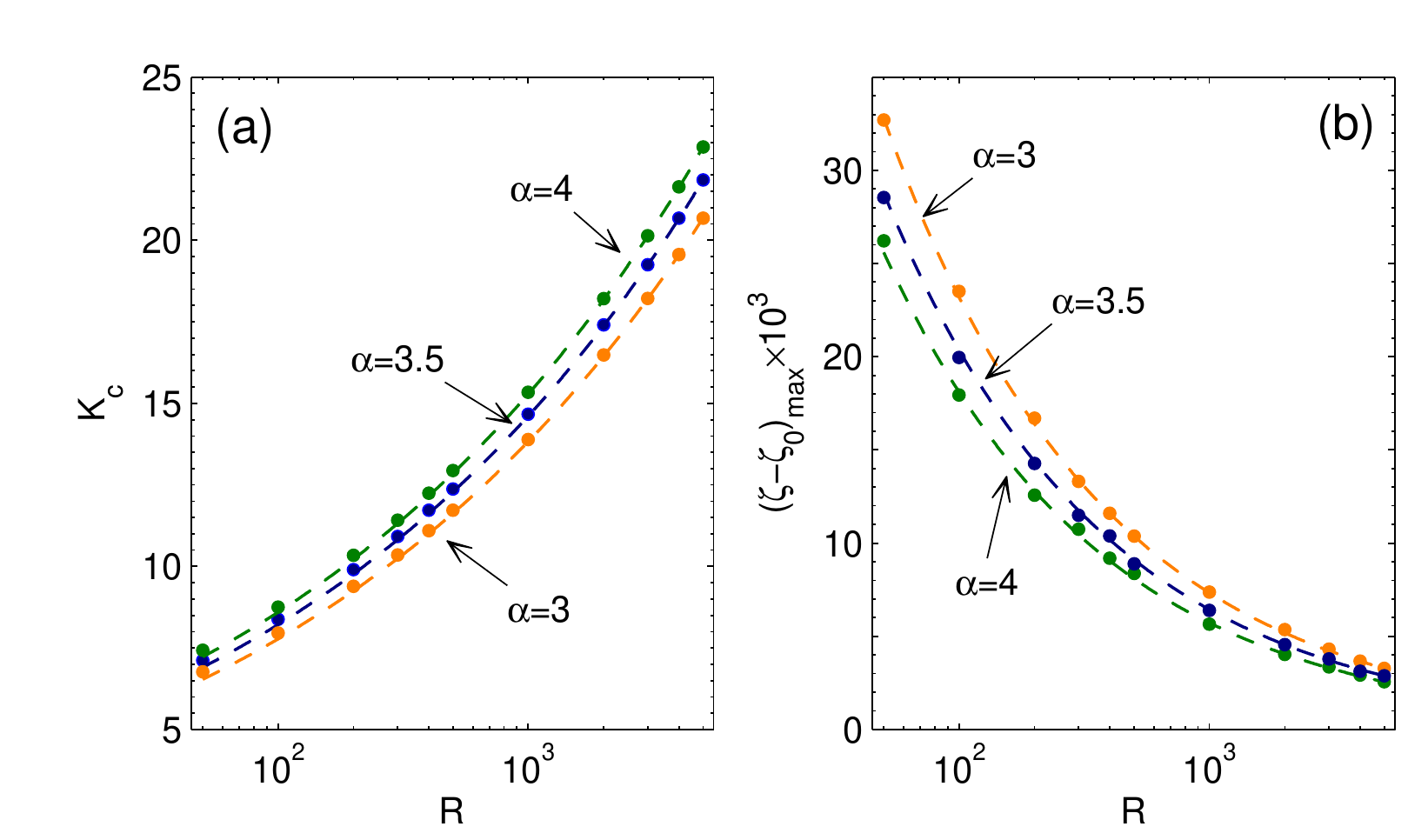}
\caption{(a) Critical parameter $K_c$ as a function of $R$ for three different values of $\alpha$. The circles are the points of actual calculation and the dashed curves demonstrate expected $R^{1/4}$ scaling. (b) Maximal deviation due to the dispersion pressure in the critical case. The dashed curves show that $R^{-1/2}$ scaling agrees well with the numerical results.\label{fig:Kc_al}}
\end{figure}

Figure \ref{fig:Kc_al}(a) shows how the critical parameter $K_c$ depends on the ratio $R=h/d_0$. When $R$ increases the force operates in a relatively short range $\xi\lesssim R^{-1/2}$ and one has to apply a larger relative force to heal the crack. The maximum of the deviation $\zeta-\zeta_0$ in the critical case is shown in Fig.$\:$\ref{fig:Kc_al}(b). It decreases roughly from $3\times 10^{-2}$ to $3\times 10^{-3}$ while $R$ increases from 50 to 5000.

To understand the results qualitatively let us approximate the pressure by a stepwise function, which is nonzero in a short lateral distance domain $0<\xi<\xi_0$. This problem can be exactly solved analytically and the detailed solution is presented in supplementary material. We provide here a simplified version of the model that is able to explain qualitatively the main features of the numerical solution. In this model the dispersion pressure is changed by the function $P(x,u)=P_C\theta(\xi_0-\xi)$, where $\theta(\xi)$ is the Heaviside step function. We take here $\xi_0=1/\sqrt{3R}$ that corresponds to the lateral distance where the pressure is reduced for about one order of magnitude: $\zeta_0(\xi_0)=1-1/R$. An approximate analytical solution of the problem (\ref{eq:beam_less}) is
\begin{equation}\label{eq:step_sol}
  \zeta-\zeta_0=\frac{K^4}{24}\left\{\begin{array}{c}
  \xi^2(\xi^2-4\xi_0\xi+6\xi_0^2),\  0<\xi<\xi_0 \\
  \xi_0^3(4\xi^3-8\xi^2+4\xi-\xi_0),\ \xi_0<\xi<1.
  \end{array}
  \right.
\end{equation}
The solution is approximate in the sense that all the coefficients in (\ref{eq:step_sol}) are given in the leading order in $\xi_0$. Because of this approximation the second and third derivatives at $\xi=\xi_0$ are discontinuous but it does not play role for what follows. Maximum of the function $\zeta-\zeta_0$ is reached at $\xi=1/3$ in agreement with the numerical solutions. The second derivative at $\xi=0$ is
\begin{equation}\label{eq:deriv_0}
  \zeta^{\prime\prime}(0)= -6+K^4\xi_0^2/2.
\end{equation}
The critical parameter is the value of $K$ for which this derivative is equal to zero; it gives $K_c=(12/\xi_0^2)^{1/4}$. Taking $\xi_0= 1/\sqrt{3R}$ we find $K_c=7.75$ for $R=100$ that is in a reasonable agreement with the value 8.39 found numerically. In the advanced variant of the model we found $\xi_0\approx\sqrt{2/3R}$ and $K_c=8.42$ (see supplementary material). Thus, we expect that $K_c$ scales asymptotically as $R^{1/4}$. This expectation fits nicely the numerical results in Fig.$\:$\ref{fig:Kc_al}(a).

The maximal value of the difference $\zeta-\zeta_0$ is realized at $\xi=1/3$. In the stepwise approximation this value is $(\zeta-\zeta_0)_{max}=2K^4\xi_0^3/81$. Using this expression we find for $K=K_c$ that the largest contribution of the dispersion pressure to the beam shape scales as $R^{-1/2}$. This scaling also works well as one can see in Fig.$\:$\ref{fig:Kc_al}(b). In dimensional terms the largest contribution of the dispersion pressure to the beam shape $u-u_0$ behaves as $(d_0h)^{1/2}$, where $u_0$ in the shape of the beam without the dispersion force.

Consider now how the energy of the deformed beam (\ref{eq:energy}) depends on the dispersion pressure. This behavior is important to relate the adhesion energy $\Gamma$ to the crack length $s$. In terms of the normalized variables the energy can be presented as (see details in supplementary material)
\begin{equation}\label{eq:energy_1}
  U=hwP_C\left(\frac{Et^3h}{12P_C}\right)^{1/4}(W_{el}+W_F),
\end{equation}
where $W_{el}$ and $W_F$ are dimensionless functionals of $\zeta$. The first one, $W_{el}$, is associated with the elastic energy of the beam. Using integration by parts it can be presented in the form
\begin{equation}\label{eq:Wel_def}
  W_{el}=\frac{\zeta^{\prime\prime\prime}(0)}{2K^3}+
  \frac{K}{2}\int_0^1d\xi\frac{\zeta}{(1+R-R\zeta)^{\alpha}}.
\end{equation}
The second one, $W_F$, is associated with the work done by the external force
\begin{equation}\label{eq:WF_def}
  W_F=-K\int_0^1d\xi\int_0^{\zeta}\frac{d\eta}{(1+R-R\eta)^{\alpha}}.
\end{equation}
Using the stepwise force model it is easy to estimate different contributions to the dimensionless energy $W=W_{el}+W_F$. The third derivative that enters Eq.$\:$(\ref{eq:Wel_def}) is proportional to the shear force at $\xi=0$ and can be found from Eq.$\:$(\ref{eq:step_sol}) as $\zeta^{\prime\prime\prime}(0)=12-K^4\xi_0$. The first term originates from the unperturbed beam but the second one is due to the dispersion interaction. When $K$ increases the shear force changes sign and becomes large in the absolute value. On the other hand, the integral term in (\ref{eq:Wel_def}) is estimated as $K\xi_0/2$. It is interesting to note that this last term cancels exactly the contribution of the dispersion pressure at $\xi=0$. All that is left is the elastic energy of the unperturbed beam $W_{el}=6/K^3$. The maximal relative correction to this expression is $\sim \xi_0\ll 1$ (see supplementary material). Actually one could expect that the elastic energy cannot change significantly due to the dispersion pressure. This is because this pressure induces a relatively small effect in the beam shape and the unperturbed beam dominates in the elastic energy.

The work done by the dispersion force is estimated from (\ref{eq:WF_def}) as $W_F\sim -K\xi_0/R$ (supplementary material). It is of the same order as the correction to $W_{el}$, which we neglected. Thus, we expect that the main contribution to the total energy of the deformed beam is $W=6/K^3$ with the relative correction $\sim\xi_0$, which scales as $R^{-1/2}$.

The discussion above shows that for qualitative analysis $W=6/K^3$ is a good approximation for the dimensionless energy. Expressing the adhesion energy as $\Gamma =-w^{-1}dU/ds$ one finds that in this approximation $\Gamma$ does not depend on the dispersion pressure:
\begin{equation}\label{eq:Gamma}
  \Gamma=-hP_C\frac{dW}{dK}\approx \frac{3Et^3h^2}{2s^4}.
\end{equation}
This is the same expression used in earlier works \cite{Boer1999,Knapp2002}.

The parameter $K$ can be expressed via the pressure $P_C$ and the adhesion energy $\Gamma$. Using Eqs.$\:$(\ref{eq:beam_less}) and (\ref{eq:Gamma}) we find
\begin{equation}\label{eq:K_express}
  K^4=18R(P_Cd_0/\Gamma).
\end{equation}
In the general case the pressure and the adhesion energy are independent parameters. Of course, adhesion includes the dispersion pressure as one of the components but not the only one. Additionally chemical interaction in the places of actual contact or locally formed water menisci can contribute to the adhesion energy. The adhesion energy is minimal if only vdW/Casimir forces are involved. These forces are omnipresent and cannot be excluded. The minimal $\Gamma$ is defined as the free energy of the vdW/Casimir interaction between the beam and the substrate separated by the average distance $d_0$. The free energy can be expressed by the Lifshitz formula \cite{Dzyaloshinskii1961} via the dielectric properties of the bodies. Note that roughness of the bodies can only increase the value of $\Gamma$ and for small $d_0$ this effect can be significant \cite{Zwol2008,Broer2011,Broer2012}.

If only the dispersion pressure contributes to the adhesion energy,  $\Gamma$ and $P_C$ have the same physical origin and, therefore, are related to each other as energy and force: $\Gamma=P_Cd_0/(\alpha-1)$. Using Eq.$\:$(\ref{eq:K_express}) we find $K=(18(\alpha-1)R)^{1/4}$. It can be compared with $K_c$ presented in Fig.$\:$\ref{fig:Kc_al}(a). For example, for $\alpha=3$ we find $K=2.45R^{1/4}$ while the critical value is just a little bit larger $K_c=2.46R^{1/4}$. It means that if the adhesion energy is defined only by the vdW/Casimir forces the adhered beam will be very close to the critical situation.

Nonlocal response of an adhered cantilever on the dispersion pressure is a  convenient property that can be used to probe the dispersion forces at small distances by measuring the effect far from the place where the force is applied. Simultaneously the pressure and the adhesion energy can be determined; the system does not suffer from the jump-to-contact problem; surface charges or contact potential do not play significant role at distances $d_0\sim 10\:$nm. One can propose a few protocols to measure the dispersion pressure using the adhered cantilever but here we shortly describe only one of them.

The force can be measured with a laser vibrometer, which is sensitive to the rate of change of the optical path. The shape of the cantilever and the unadhered length $s$ can be determined with a high precision by scanning with the laser beam along the cantilever. To make estimates we are using simple expressions for the shape (\ref{eq:zeta_0}) and (\ref{eq:step_sol}). The vibrometer signal (velocity) as a function of time $\tau$ is given by
\begin{eqnarray}\label{eq:signal}
   \nonumber S_0 &=& -6\frac{v_s h}{s}\xi(1-\xi),\ \xi=\frac{v_s \tau}{s} \\
  \Delta S &\approx & \frac{v_s\sqrt{d_0h}}{\sqrt{3}s}
  \left(\frac{P_Cd_0}{\Gamma}\right)(3\xi^2-4\xi+1),
\end{eqnarray}
where $v_s$ is the scan speed, $S_0$ is the signal from unperturbed beam, $\Delta S$ is the change of the signal due to the dispersion force, and the total signal is $S=S_0+\Delta S$. It is assumed that in Eq.$\:$(\ref{eq:signal}) $\xi>\xi_0$. The contrast is estimated as
\begin{equation}\label{eq:contrast}
  \frac{\Delta S}{S_0}\approx -\left(\frac{P_Cd_0}{\Gamma \sqrt{3R}}\right)\frac{3\xi^2-4\xi+1}{6\xi(1-\xi)}.
\end{equation}
This ratio is zero at $\xi=1/3$, where $\zeta-\zeta_0$ is maximal, but it is strongly increases for $\xi\rightarrow \xi_0$ where it is as large as $P_Cd_0/6\Gamma$. Significant increase of the contrast happens because the unperturbed shape increases quadratically while the perturbed shape increases linearly with $\xi$ for $\xi>\xi_0$. The absolute value of the signal is controlled by the scan speed and  is well measurable at $v_s>1\:$m/s. Since the largest contrast does not depend on the cantilever parameters they can be chosen in a wide range. The cantilevers can be microfabricated with a thickness of $2\:\mu$m and a length of $1\:$mm as in \cite{Knapp2002} or made of thin ($50\:\mu$m) silicon wafers with a length of $50\:$mm or so.

In conclusion, we considered the influence of the vdW/Casimir forces on an adhered cantilever. Although the forces are operative only very close to the point of contact, they influence the shape of the cantilever far from the contact at about one third of the unadhered length. The canilever can be used to measure simultaneously the dispersion forces and the adhesion energy at short separations, where the usual methods suffer from the jump-to-contact instability.

\section*{Supplementary Material}

See supplementary material for exact solution of the stepwise force model and for energy calculation.

\end{document}